\documentclass[12pt,showpacs,a4paper,prb]{revtex4}
\usepackage{amsmath}
\usepackage{amssymb}
\begin{document}

\title{Classical Weyl-Spinor approach to $U(1)$ and non-abelian local gauge theories}

\author{J.Buitrago}
\author{S. Hajjawi}
\affiliation{Department of Astrophysics of the University of La Laguna, Avenida
Francisco Sanchez, s/n, 38205, La Laguna, Tenerife, Spain}
\email{jgb@iac.es}
\email{serena_hajjawi@yahoo.es}
\date{\today}

\begin{abstract}
In a previous paper we introduced two linear spinor
equations equivalent to the Lorentz Force and stated that these equations were fairly general and could 
be applied to any force field compatible with Special Relativity. In this paper, via a lagrangian approach,
we explore this possibility and obtain classical spinor equations describing the behaviour of fermionic
particles not only under an electromagnetic field but also under Yang-Mills and Color fields.
We find a covariant derivative defined {\it along the classical trajectory} of the particle, 
which can be extended to SU(2) and SU(3) local symmetries, and obtain the Yang-Mills and Color fields 
in a new classical Weyl-spinor approach to Gauge Theories. In the  SU(3) case, 
the obtained equations which describe the behaviour of quarks under gluon fields could be 
in principle applied to the quark-gluon plasma phase existing during the first instants of the Universe.

\end{abstract} 

\maketitle

\section{INTRODUCTION}

As is well known the lagrangian of a free fermion \\
\begin{displaymath}
L=i\bar{\psi}\gamma^{\mu}\partial_{\mu}\psi-m\bar{\psi}\psi,
\end{displaymath} \\
leading to the Dirac equation is not invariant under local phase transformations of the form \\
\begin{displaymath}
\begin{array}{c}
\psi \rightarrow exp\{ie\varphi(x)\}\psi \\
\bar{\psi} \rightarrow exp\{-ie\varphi(x)\}\bar{\psi} .\\
\end{array}
\end{displaymath} \\
In order to achieve invariance it is necessary to add a term \\
\begin{displaymath}
L_{I}=e\left(\bar{\psi}\gamma^{\mu}\psi\right)A_{\mu}
\end{displaymath} \\
where the new vector-field $A_{\mu}$ must transform, coupled to the phase-factor, as \\
\begin{displaymath}
A_{\mu} \rightarrow A_{\mu}+\frac{1}{e}\partial_{\mu}\varphi,
\end{displaymath} \\
thus local gauge invariance produce the electromagnetic interaction in terms of the potential 
$A_{\mu}$. Now, in this work, we would like to ask the question of what are the consequences of 
applying the gauge principle
in the classical spinor formalism emerging from the $Geometrical$ $Principle$ introduced in 
a precedent paper, which shall be referred to as paper I \cite{buitrago}.
We shall see here how, in Weyl-Spinor language, local 
gauge invariance emerges at a deeper level producing the $4-rank$ spinor describing the electromagnetic
field in terms of its physical components (i.e. the electric and the magnectic field 
strengths). To start with this, we note that, in its more simple form, gauge invariance 
is already
present in the spinor representation of a photon (see paper I), given by a null hermitian spinor
(we use Penrose standard notation \cite{penrose}) as\\
\begin{displaymath}
\pi^{A}\bar{\pi}^{A'},
\end{displaymath} \\
which is manifestly invariant under the transformation \\
\begin{displaymath}
\begin{array}{c}
\pi^{A} \rightarrow exp\{ie\varphi(x)\}\pi^{A} \\
\bar{\pi}^{A'} \rightarrow exp\{-ie\varphi(x)\}\bar{\pi}^{A'}. \\
\end{array}
\end{displaymath} \\
\section{THE SPINORIAL LAGRANGIAN}
In paper I we obtained the henceforth called $master$ $equations$  for a charged particle of mass $m$
and four-momentum \\
\begin{displaymath}
p^{AA'}=\frac{1}{\sqrt{2}}\left[\pi^{A}\bar{\pi}^{A'}+\eta^{A}\bar{\eta}^{A'}\right]
\end{displaymath} \\
in the presence of an electromagnetic field described by \\
\begin{displaymath}
F_{AA'BB'}=\epsilon_{AB}\bar{\phi}_{A'B'}+\epsilon_{A'B'}\phi_{AB}.
\end{displaymath} \\
The equations being:  \\
\begin{equation} \label{1.1}
\begin{array}{c}
\dot{\eta}_{A}=-\frac{e}{m}\phi_{AB}\eta^{B} \\
\dot{\pi}_{A}=-\frac{e}{m}\phi_{AB}\pi^{B},\\
\end{array}
\end{equation} \\
where the dot means derivative with respect to proper time $\tau$
(similar expressions for the complex-conjugate spinors hold). These equation were obtained by applying the
$Geometrical$ $Principle$ to the elements of $S(2,C)$ (i,e., to the individual spinors).
However, to give a more complete picture we would like here to re-derive 
them in analogy with the standard theory, from the $Gauge$ 
$Principle$. To this end, we start by defining the free action for a particle of mass 
$m$ as \\
\begin{equation} \label{action}
S_{f} =\frac{1}{m} \int d\tau \dot{\eta}^{A}\pi_{A}
\end{equation} \\
so that the $minimal$ $action$ $principle$, when applied with respect to $\eta^{A}$ \\ 
\begin{displaymath}
\begin{array}{c}
\delta S_{f}=0 =\frac{1}{m}\int_{\tau_{1}}^{\tau_{2}} d\tau \delta\dot{\eta}^{A}\pi_{A}= \\
\frac{1}{m}\int_{\tau_{1}}^{\tau_{2}} d\tau \left(\frac{d}{d\tau}\delta\eta^{A}\right)\pi_{A}= \\
\frac{1}{m}\int_{\tau_{1}}^{\tau_{2}} d\tau \frac{d}{d\tau}(\delta\eta^{A}\pi_{A})-
\frac{1}{m}\int_{\tau_{1}}^{\tau_{2}} d\tau \delta\eta^{A}\dot{\pi}_{A}, \\
\end{array}
\end{displaymath} \\ 
leads trivially to $\dot{\pi}_{A}=0$, since the first integral on the third line of the above equality 
is null. It is also possible to write $S_{f}=\frac{1}{m}\int d\tau
\dot{\pi}^{A}\eta_{A}$ and consider variations of
the path with respect to $\pi^{A}$ instead of to $\eta^{A}$. Both possibilities are fully equivalent,
since
$\eta^{A},\pi^{A}$ are constrained by the condition that $\eta^{A}\pi_{A}$ is a constant of motion (see Eq. 7 in paper I). Moreover, 
once we have fixed the components of $\eta^{A}$ those of $\pi^{A}$ are, up to a constant phase factor, 
completely determined and viceversa. 
The lagrangian corresponding to the action (\ref{action}) is given by \\ 
%
%
%
\begin{equation} \label{1.2}
L_{f}=\frac{1}{m}\dot{\eta}^{A}\pi_{A}
\end{equation} \\
and, the Euler-Lagrange equations \\

\begin{equation} \label{1.3}
\frac{d}{d\tau}\frac{\partial L}{\partial \dot{\eta}^{A}}-\frac{\partial L}{\partial \eta^{A}}=0 
\end{equation} \\
when applied to it lead of course to \\
\begin{equation} \label{1.5}
\dot{\pi}_{A}=0 \Longrightarrow \pi_{A}=const. \\
\end{equation} \\
It follows then that $\dot{\eta}^{A}$ must also be null, since $\dot{\eta}^{A}\pi_{A}=\dot{\pi}^{A}\eta_{A}$
(this is due to the condition of conserved rest-mass in spinorial representation, given by  
$\eta^{A}\pi_{A}=constant$).
We now explore the consequences of local 
gauge invariance. Let us 
apply the following transformations ($e$ is the electric charge) {\it along the classical trajectory
of the particle}\\
\begin{equation} \label{1.6}
\begin{array}{c}
\eta_{A} \rightarrow exp\{ie\frac{\varphi(\tau)}{2}\}\eta_{A} \\
\pi_{A} \rightarrow exp\{ie\frac{\psi(\tau)}{2}\}\pi_{A} \\
\end{array}
\end{equation} \\
Now, since the condition\\
\begin{equation} \label{1.7}
\frac{d}{d\tau}(\eta_{A}\pi^{A}) \rightarrow \frac{d}{d\tau}\left(exp\{ie\frac{\varphi+\psi}{2}\}
\eta_{A}\pi^{A}\right)=\left[\frac{ie}{2}(\dot{\varphi}+\dot{\psi})\eta_{A}\pi^{A}+
\frac{d(\eta_{A}\pi^{A})}{d\tau}\right]exp\{ie\frac{\varphi+\psi}{2}\}=0
\end{equation} \\
must hold, $\varphi$ and $\psi$ should be related by\\
\begin{equation} \label{1.8}
\dot{\varphi}=-\dot{\psi} \Longrightarrow \varphi=-\psi + const,
\end{equation} \\
so, up to a constant phase-factor, the phase shifts of $\eta_{A}$ and $\pi_{A}$ must be
opposite: \\
\begin{equation} \label{1.9}
\begin{array}{c}
\eta_{A} \rightarrow exp\{ie\frac{\varphi(\tau)}{2}\}\eta_{A} \\
\pi_{A} \rightarrow exp\{i\frac{k}{2}-ie\frac{\varphi(\tau)}{2}\}\pi_{A}. \\
\end{array}
\end{equation} \\
With this transformation, the free-lagrangian reads now\\
\begin{equation} \label{1.10}
L_{f} \rightarrow exp\{i\frac{k}{2}\}\frac{1}{m}\left(\frac{ie}{2}\dot{\varphi}\eta^{A}\pi_{A}+
\dot{\eta}^{A}\pi_{A}\right)
\end{equation} \\
and, the additional term (neglecting the constant factor) appearing in it can be written as \\
\begin{equation} \label{1.11}
\frac{1}{2m}ie\dot{\varphi}\eta^{A}\pi_{A}=
\frac{ie}{2m}\dot{\varphi}\epsilon_{AB}\eta^{B}\pi^{A}.
\end{equation} \\
It is  easy to see that, if we add an interaction term of the form \\
\begin{equation} \label{1.12}
-\frac{e}{m^{2}}\phi_{AB}\eta^{B}\pi^{A}
\end{equation} \\
and impose the condition to the new field $\phi_{AB}$ of transforming, under local phase
transformations, as \footnote{The validity of transformation (13) is consequence of the following theorem applied to valence-2 spinors (see Steward. J.{\it Advanced General Relativity}. 1991 Cambridge Univ. Press. Page 69): ``Any spinor $\tau_{A...F}$ is the sum of the totally symmetric spinor $\tau_{(A...F)}$ and (outer) products of $\epsilon 's$ with totally symmetric spinors of lower valence"}\\
\begin{equation} \label{1.13}
\phi_{AB} \rightarrow \phi_{AB}+im\frac{\dot{\varphi}}{2}\epsilon_{AB},
\end{equation} \\
then, the new lagrangian \\
\begin{equation} \label{1.14}
L=\frac{1}{m}\dot{\eta}^{A}\pi_{A} -
\frac{e}{m^{2}}\phi_{AB}\eta^{B}\pi^{A},
\end{equation} \\
is invariant under $U(1)$ local-phase transformations. The transformation 
that holds
for the conjugate second-rank spinor   $\bar{\phi}^{A'B'}$,  is
given by \\
\begin{equation} \label{1.15}
\bar{\phi}_{A'B'} \rightarrow \bar{\phi}_{A'B'}-im\frac{\dot{\varphi}}{2}\epsilon_{A'B'}.
\end{equation} \\
This kind of transformations leave however
invariant the associated
four-rank spinor of the Maxwell field strength \\
\begin{displaymath}
F_{AA'BB'}=\epsilon_{AB}\bar{\phi}_{A'B'}+\epsilon_{A'B'}\phi_{AB}
\end{displaymath} \\
since, according to (\ref{1.13}),(\ref{1.15}), transforms as \\
\begin{equation} \label{1.16}
\epsilon_{AB}\bar{\phi}_{A'B'}+\epsilon_{A'B'}\phi_{AB} \rightarrow
\epsilon_{AB}\bar{\phi}_{A'B'}+\epsilon_{A'B'}\phi_{AB} +im\frac{\dot{\varphi}}{2}\epsilon_{A'B'}\epsilon_{AB}-
im\frac{\dot{\varphi}}{2}\epsilon_{AB}\epsilon_{A'B'},
\end{equation} \\
so the physical components of the electromagnetic field are unchanged under $U(1)$
transformations, just as we should expect. On the other hand,
in analogy with Classical Mechanics, from the lagrangian (\ref{1.14}),
we may also define a hamiltonian by setting \\
\begin{equation} \label{1.17}
H=\frac{\partial L}{\partial \dot{\eta}^{A}}\dot{\eta}^{A}- L=
\frac{\partial L}{\partial \dot{\eta}_{A}}\dot{\eta}_{A}- L=
\frac{e}{m^{2}}\phi_{AB}\eta^{B}\pi^{A},
\end{equation} \\
since \\
\begin{equation} \label{1.19}
\frac{\partial L}{\partial \dot{\eta}^{A}}=\frac{1}{m}\pi_{A} \equiv \hat{\pi}_{A}, 
\end{equation} \\
so the conjugate momentum is $p_{A}=\hat{\pi}_{A}$ (equivalently $p^{A}=-\hat{\pi}^{A}$). The 
Hamilton equations \\
\begin{equation} \label{1.21}
\frac{\partial H}{\partial \hat{\pi}_{A}}=\dot{\eta}^{A}
\end{equation}
\begin{equation} \label{1.22}
\frac{\partial H}{\partial \eta^{A}}=-\dot{\hat{\pi}}_{A} 
\end{equation} \\
then lead to the $master$ $equations$: \\
\begin{equation} \label{1.23}
\dot{\eta}^{A}=\frac{e}{m}\phi^{AB}\eta_{B}
\end{equation} 
\begin{equation} \label{1.24}
\dot{\pi}_{A}=-\frac{e}{m}\phi_{AB}\pi^{B}.
\end{equation} \\
Although the spinor hamiltonian given by (\ref{1.17}) is entirely classical it can also give
discrete values for the energy in appropiate situations. However, as is well-known, the 
number of these cases is severely constrained by the requirement of Lorentz Covariance.
In order to show that equations (\ref{1.1}) describe $1/2$-spin particles (see also paper I)
we shall consider a charged particle of mass $m$ in a constant magnetic field. In the rest frame of the particle  
the only non-null components of the field spinor $\phi_{AB}$ are (we take the magnetic field 
$\vec{B}$ along the $z$-axis) \\
\begin{displaymath}
\phi_{01}=\phi_{10}=-\frac{i}{2}B_{0}
\end{displaymath} \\
and the individual spinor solutions to the master equations are then\\
\begin{displaymath}
\pi^{A}(\tau)=\sqrt{m}e^{\pm i\frac{\pi}{2}}\left(\begin{array}{c}
e^{-ie\frac{B_{0}}{2m}\tau} \\
0 \\
\end{array}\right),\quad \eta^{A}(\tau)=\sqrt{m}\left(\begin{array}{c}
0 \\
e^{ie\frac{B_{0}}{2m}\tau} \\
\end{array}\right)
\end{displaymath} \\
where the global phase factor for the spinor $\eta^{A}(\tau)$
has been set to unity (this is always possible since solutions are completely 
determined up to a constant factor). By substitution in 
the hamiltonian (\ref{1.17}) the values of the energies are (the global phase factors of $\pi^{A}$ and $\eta^{A}$ 
respectively, in the rest frame, are constrained by the condition that 
their difference must equal $i\frac{\pi}{2}$ times 
an odd integer number \cite{buitrago}, so two different values of energy are in
this case allowed )\\
\begin{displaymath}
E=\pm\frac{e}{2m}B_{0}.
\end{displaymath} \\
\section{Covariant Derivative}

For further generalizations to non abelian symmetries it will be useful to define a covariant derivative. To this end, we start
with the free Lagrangian \\
\begin{displaymath}
L_{f}=\frac{1}{m}\dot{\eta}^{A}\pi_{A}
\end{displaymath} \\
and define a covariant derivative along the {\it{classical trajectory of the particle}} (i.e. a total 
derivative) in the following way \\
\begin{equation} \label{covariant}
\frac{D}{d\tau}\eta^{A}\equiv \frac{d\eta^{A}}{d\tau}-\frac{e}{m}\phi^{AB}\eta_{B}.
\end{equation} \\
Now it is easy to check that the interacting Lagrangian obtained in the preceeding $section$ can be obtained by
merely replacing the ordinary derivative by the covariant one: \\
\begin{displaymath}
\begin{array}{c}
L_{int}=\frac{1}{m}\left(\frac{D}{d\tau}\eta^{A}\right)\pi_{A} \\
\frac{1}{m}(\dot{\eta}^{A}-\frac{e}{m}\phi^{AB}\eta_{B})\pi_{A}= \\
\frac{1}{m}\dot{\eta}^{A}\pi_{A}-\frac{e}{m^{2}}\phi^{AB}\eta_{B}\pi_{A}. \\
\end{array}
\end{displaymath} \\
But we would like to show that the covariant derivative (\ref{covariant}) in this way defined is also
a $gauge$ derivative which in the $U(1)$ context means that, if under a local change of phase
(along the classical trajectory)\\
\begin{displaymath} 
\eta'_{A}=exp\{ie\frac{\varphi(\tau)}{2}\}\eta_{A}
\end{displaymath} \\
then \\
\begin{displaymath}
\frac{D'}{d\tau}\eta'_{A}=exp\{ie\frac{\varphi(\tau)}{2}\}\frac{D}{d\tau}\eta_{A}.
\end{displaymath} \\
\underline{$Proof$} \\
\begin{displaymath}
\begin{array}{c}
\frac{D'}{d\tau}\eta'_{A}=\frac{d}{d\tau}
\left(e^{ie\frac{\varphi(\tau)}{2}}\eta_{A}\right)
+\frac{e}{2m}\phi'_{AB}\epsilon^{BC}
e^{ie\frac{\varphi(\tau)}{2}}\eta_{C}= \\
ie\frac{\dot{\varphi}}{2}e^{ie\frac{\varphi(\tau)}{2}}\eta_{A}+e^{ie\frac{\varphi(\tau)}{2}}\dot{\eta}_{A}+
\frac{e}{2m}\left(\phi_{AB}+im\dot{\varphi}\epsilon_{AB}\right)\epsilon^{BC}e^{ie\frac{\varphi(\tau)}{2}}\eta_{C} \\
\end{array}
\end{displaymath} \\
where we have made the replacement \\
\begin{displaymath}
\phi'_{AB}=\phi_{AB}+im\frac{\dot{\varphi}}{2}\epsilon_{AB}
\end{displaymath} \\
which, as previously obtained, is the transformation rule of $\phi_{AB}$ under a local phase change
so the $\dot{\varphi}$ terms cancel out and we get \\
\begin{displaymath}
\frac{D'}{d\tau}\eta'_{A}=e^{ie\frac{\varphi(\tau)}{2}}\dot{\eta}_{A}+\frac{e}{2m}\phi_{AB}\epsilon^{BC}e^
{ie\frac{\varphi(\tau)}{2}}\eta_{C}=e^{ie\frac{\varphi(\tau)}{2}}\frac{D}{d\tau}\eta_{A}.
\end{displaymath} \\
\section{Non Abelian Symmetries and Classical Equations of Motion}
In tensor language, purely classical theories in which is still possible to speak about particle
trajectories parametrized by proper time and consistent with special relativity end with 
electrodynamics. However, in Weyl-spinor language, the classical formalism developed in 
the precedent section can be extended to describe other interactions as well. Although our aim
is to obtain classical equations of motion for quarks and the color fields, for the sake of clarity
we shall first consider the local $SU(2)$ symmetry leading to the Yang-Mills fields. Then, the extension
to $SU(3)$ is inmediate. To go over this subject first consider the following identities (we adopt the 
convention that the scalar $\eta^{A}\pi_{A}$ is real and equals the rest-mass $m$ of the
particle under consideration)\\
\begin{displaymath}
\begin{array}{c}
p^{AA'}\pi_{A}=\frac{m}{\sqrt{2}}\bar{\eta}^{A'} \\ 
p_{AA'}\bar{\eta}^{A'}=\frac{m}{\sqrt{2}}\pi_{A} \\
\end{array}
\end{displaymath} \\
which can be summarized into a single expression\\
\begin{equation} \label{dirac1}
\gamma^{\mu} p_{\mu}\psi=m\psi
\end{equation} \\
by means of the bispinor (if we were considering both $\pi^{A}$ and $\eta^{A}$ as being 
distributions over the momentum-space we should speak of $\psi$ as a Dirac bispinor. However this is not the case: 
$\eta^{A},\pi^{A}$ are here defined along the classical trajectory of the particle) \\
\begin{displaymath}
\psi=\frac{1}{\sqrt{2}}\left(\begin{array}{c}
\pi_{A} \\
\bar{\eta}^{A'} \\
\end{array}\right)
\end{displaymath} \\
being $\gamma^{\mu}$ the Weyl representation of the well-known $gamma$ matrices. Then the Lorentz scalar \\
\begin{displaymath}
\bar{\psi}\psi=\psi^{\dag}\gamma^{0}\psi=\frac{1}{2}\left(\begin{array}{cc}
\bar{\pi}_{A'} & \eta^{A} \\
\end{array}\right)
\left(\begin{array}{cc}
0 & 1 \\
1 & 0 \\
\end{array}\right)\left(\begin{array}{c}
\pi_{A} \\
\bar{\eta}^{A'} \\
\end{array}\right)=\eta^{A}\pi_{A}=m
\end{displaymath} \\
is invariant under local $U(1)$ transformations \\
\begin{displaymath} 
\psi\rightarrow \psi'=exp\{ie\frac{\varphi(\tau)}{2}\}\psi. 
\end{displaymath} \\
A $SU(2)$ transformation of the form  \\
\begin{equation} \label{trans}
\psi\rightarrow \psi'=exp\{i\frac{g}{2}\vec{\tau}\cdot\vec{\varphi}(\tau)\}\psi
\end{equation} \\
($\vec{\tau}$ is the iso-vector containing the three Pauli matrices) requires however
to act upon a two-component bispinor \\
\begin{displaymath}
\psi=\left(\begin{array}{c}
\psi_{1} \\
\psi_{2} \\
\end{array}\right)
\end{displaymath} \\
made up from the tensorial product \\
\begin{equation} \label{tensorial}
\psi=\frac{1}{\sqrt{2}}\left(\begin{array}{c}
\pi_{A} \\
\bar{\eta}^{A'} \\
\end{array}\right)\otimes (\alpha\bold{\hat{\xi}}+\beta\bold{\hat{\eta}})=
\frac{1}{\sqrt{2}}\left(\begin{array}{c}
\alpha\left(\begin{array}{c} 
\pi_{A} \\
\bar{\eta}^{A'} \\
\end{array}\right) \\
{ } \\
\beta\left(\begin{array}{c}
\pi_{A} \\
\bar{\eta}^{A'} \\
\end{array}\right) \\
\end{array}\right) =\frac{1}{\sqrt{2}}\left(\begin{array}{c}
\left(\begin{array}{c}
\pi_{1A} \\
\bar{\eta}^{A'}_{1} \\
\end{array}\right) \\
{ } \\
\left(\begin{array}{c}
\pi_{2A} \\
\bar{\eta}^{A'}_{2} \\
\end{array}\right) \\
\end{array}\right)\equiv \left(\begin{array}{c}
\psi_{1} \\
\psi_{2} \\
\end{array}\right)
\end{equation} \\
where $\{\bold{\hat{\xi}},\bold{\hat{\eta}}\}$ is some orthonormal basis in the new $2$-complex-dimensional vector 
space introduced. Note that via (\ref{tensorial}) we have defined a new set of four spinors 
(barred quantities mean complex conjugate) \\
\begin{equation} \label{spinors}
\begin{array}{c}
\pi_{1A}=\alpha\pi_{A} \\
\pi_{2A}=\beta\pi_{A} \\
\eta^{A}_{1}=\bar{\alpha}\eta^{A} \\
\eta^{A}_{2}=\bar{\beta}\eta^{A} \\
\end{array}
\end{equation} \\
which englobe the coefficients $\alpha,\beta$. A transformation on $\{\alpha,\beta\}$ then
induces a transformation on $\{\pi^{A}_{a},\eta^{A}_{a}\}$ ($a=1,2$).
According to (\ref{tensorial}), (\ref{dirac1}) should now read \\
\begin{equation} \label{dirac2}
\left(\begin{array}{cc}
\gamma^{\mu}p_{\mu} & 0 \\
0 & \gamma^{\mu}p_{\mu} \\
\end{array}\right) \left(\begin{array}{c}
\psi_{1} \\
\psi_{2} \\
\end{array}\right)=m\left(\begin{array}{c}
\psi_{1} \\
\psi_{2} \\
\end{array}\right).
\end{equation} \\
Actually, under (\ref{trans}) (taken as infinitesimal) the coefficients $\alpha,\beta$ 
transform as \\
\begin{displaymath}
\left(\begin{array}{c}
\alpha \\
\beta \\
\end{array}\right)\rightarrow \left(\begin{array}{c}
\alpha \\
\beta \\
\end{array}\right)+i\frac{g}{2}\left(\begin{array}{cc}
\varphi_{3} & \varphi_{1}-i\varphi_{2} \\
\varphi_{1}+i\varphi_{2} & -\varphi_{3} \\
\end{array}\right)\left(\begin{array}{c}
\alpha \\
\beta \\
\end{array}\right). 
\end{displaymath} \\
and the induced transformation
on $\{\pi_{aA},\eta^{A}_{a}\}$ is then given by \\
\begin{equation} \label{trans1}
\left(\begin{array}{c}
\pi_{1A} \\
\pi_{2A} \\
\end{array}\right)\rightarrow
\left(\begin{array}{c}
\pi_{1A} \\
\pi_{2A} \\
\end{array}\right)+i\frac{g}{2}\left(\begin{array}{cc}
\varphi_{3} & \varphi_{1}-i\varphi_{2} \\
\varphi_{1}+i\varphi_{2} & -\varphi_{3} \\
\end{array}\right)\left(\begin{array}{c}
\pi_{1A} \\
\pi_{2A} \\
\end{array}\right)
\end{equation}
\begin{equation} \label{trans2}
\left(\begin{array}{c}
\eta^{A}_{1} \\
\eta^{A}_{2} \\
\end{array}\right)\rightarrow
\left(\begin{array}{c}
\eta^{A}_{1} \\
\eta^{A}_{2} \\
\end{array}\right)-i\frac{g}{2}\left(\begin{array}{cc}
\varphi_{3} & \varphi_{1}+i\varphi_{2} \\
\varphi_{1}-i\varphi_{2} & -\varphi_{3} \\
\end{array}\right)\left(\begin{array}{c}
\eta^{A}_{1} \\
\eta^{A}_{2} \\
\end{array}\right),
\end{equation} \\
leaving the scalar (Einstein's summation convention over lower case indexes is understood) \\
\begin{displaymath}
\bar{\psi}\psi=\psi^{\dag}\gamma^{0}\psi=Re\{\eta^{A}_{1}\pi_{1A}+\eta^{2}_{A}\pi_{2A}\}=
(\vert\alpha\vert^{2}+\vert\beta\vert^{2})\eta^{A}\pi_{A}=\eta^{A}_{a}\pi_{aA}
\end{displaymath} \\
invariant due to the unitary character of (\ref{trans}). 
If we want this scalar to equal the rest-mass $m$ as in the $U(1)$ case, we must impose \\
\begin{displaymath}
\vert\alpha\vert^{2}+\vert\beta\vert^{2}=1.
\end{displaymath} \\
We now relate the set of spinors $\{\pi^{A}_{a},\eta^{A}_{a}\}$ in (\ref{spinors}) 
to a fermionic particle of mass $m$ and four-momentum ($a,b=1,2$) \\
\begin{displaymath}
p^{AA'}_{ab}=\frac{1}{\sqrt{2}}\left(\pi^{A}_{a}\bar{\pi}^{A'}_{b}+
\eta^{A}_{a}\bar{\eta}^{A'}_{b}\right),
\end{displaymath} \\
consistent with \\
\begin{displaymath}
p^{AA'}_{ab}p_{abAA'}=m^{2}.
\end{displaymath}\\
We are then concerned with spinorial gauge-fields leading to equations of motion for 
charged particles which remain invariant under (\ref{trans}) and preserve,
along the trajectory, the scalar quantity $\eta^{A}_{a}\pi_{aA}$ so that \\
\begin{displaymath}
\frac{d}{d\tau}(\eta^{A}_{a}\pi_{aA})=0.
\end{displaymath} \\
Let us now explore the consequences of $SU(2)$ local phase transformations along the classical path
of the particle. Note that the relations (\ref{trans1}),(\ref{trans2}) 
in matrix-form are equivalent to \\
\begin{displaymath}
\pi_{aA} \rightarrow \pi'_{aA}=[exp\{\frac{i}{2}g\vec{\tau}\cdot\vec{\varphi}(\tau)\}]_{ab}\pi_{bA}
\end{displaymath}
\begin{displaymath}
\eta_{aA} \rightarrow \eta'_{aA}=[exp\{-\frac{i}{2}g\vec{\tau}^{*}\cdot\vec{\varphi}(\tau)\}]_{ab}\eta_{bA}, 
\end{displaymath}\\
where $\vec{\tau}^{*}$ is the transpose iso-vector and $g$, as usual, a coupling constant.
The covariant derivative should then be of the form\\
\begin{equation}\label{covariant1}
\frac{D\eta_{aA}}{d\tau}\equiv \frac{d\eta_{aA}}{d\tau}+[\frac{g}{m}\vec{\tau}^{*}\cdot\vec{\chi}_{AB}]_{ab}\eta^{B}_{b},
\end{equation} \\
where $\vec{\chi}_{AB}$ is an iso-spinor-vector that should transform according to the adjoint three dimensional 
representation of $SU(2)$. In order to find out how  $\vec\chi_{AB}$  transform we note that from the 
gauge derivative condition \\
\begin{displaymath} 
\frac{D'}{d\tau}\eta'_{aA}=[exp\{-\frac{i}{2}g\vec{\tau}^{*}\cdot\vec{\varphi}(\tau)\}]_{ab}\frac{D}{d\tau}\eta_{bA}
\end{displaymath} \\
and, considering an infinitesimal transformation with parameter $\vec{\alpha}(\tau)$, the left-hand side of the above equation
is \\
\begin{equation} \label{covariant2}
\begin{array}{c}
\frac{D'}{d\tau}\eta'_{aA}=\frac{D'}{d\tau}\left[1-i\frac{g}{2}\vec{\tau}^{*}\cdot\vec{\alpha}(\tau)\right]_{ab}\eta_{bA}=
\frac{D'}{d\tau}\left[\eta_{aA}-i\frac{g}{2}\left(\vec{\tau}^{*}\cdot\vec{\alpha}(\tau)\right)_{ab}\eta_{bA}\right]= \\
\frac{d\eta_{aA}}{d\tau}+(\frac{g}{m}\vec{\tau}^{*}\cdot\vec{\chi}'_{AB})_{ab}\eta_{b}^{B}+
\frac{d}{d\tau}\left[\left(-\frac{ig}{2}\vec{\tau}^{*}\cdot\vec{\alpha}(\tau)\right)_{ab}\eta_{bA}\right]+
(\frac{g}{m}\vec{\tau}^{*}\cdot\vec{\chi}'_{AB})_{ab}\left[\frac{-ig}{2}\vec{\tau}^{*}\cdot\vec{\alpha}(\tau)\right]_{bc}\eta_{cA} \\
\end{array}
\end{equation} \\
while the right-hand side equals to \\
\begin{displaymath}
\dot{\eta}_{aA}+(\frac{g}{m}\vec{\tau}^{*}\cdot\vec{\chi}_{AB})_{ab}\eta_{b}^{B}-(\frac{i}{2}g\vec{\tau}^{*}\cdot\vec{\alpha}(\tau))
_{ab}\dot{\eta}_{bA}-(\frac{i}{2}g\vec{\tau}^{*}\cdot\vec{\alpha}(\tau))_{ab}(\frac{g}{m}\vec{\tau}^{*}\cdot\vec{\chi}_{AB})_{bc}\eta_{c}^{B}
\end{displaymath} \\
Since $\vec\alpha(\tau)$ is infinitesimal we assume \\
\begin{displaymath}
\vec\chi_{AB}\rightarrow \vec\chi_{AB}+\delta\vec\chi_{AB}
\end{displaymath} \\
where $\delta\vec\chi_{AB}$ is small. By substitution in (\ref{covariant2}), after some lengthy calculations
and using the well-known relation \\
\begin{displaymath}
(\vec{a}\cdot\vec{\tau}^{*})(\vec{b}\cdot\vec{\tau}^{*})=\vec{a}\cdot\vec{b}+
i(\vec{a}\times\vec{b})\cdot\vec{\tau}^{*},
\end{displaymath} \\
we find \\
\begin{equation} \label{covariant3}
\vec{\chi}'_{AB}=\vec{\chi}_{AB}-im\dot{\vec{\alpha}}\epsilon_{AB}-g(\vec{\alpha}\times\vec{\chi}_{AB}).
\end{equation} \\
To find the spinor-vector fields $\vec F_{AA'BB'}$ we note that, since every component must be skew, it is
always possible to write \\
\begin{equation} \label{covariant4}
\vec F_{AA'BB'}=\vec\chi_{AB}\epsilon_{A'B'}+\vec{\bar{\chi}}_{A'B'}\epsilon_{AB} 
\end{equation} \\
where $\vec{\bar{\chi}}_{A'B'}$ is also symmetric and transforms as \\
\begin{displaymath}
\vec{\bar{\chi}'}_{A'B'}=\vec{\bar{\chi}}_{A'B'}+im\vec{\dot\alpha}\epsilon_{A'B'}-g(\vec\alpha\times\vec{\bar{\chi}}_{A'B'}).
\end{displaymath} \\
In turn, $\vec F_{AA'BB'}$ should transform as an iso-vector in the following way\\
\begin{equation} \label{covariant5}
\vec{F}'_{AA'BB'}=\vec{F}_{AA'BB'}-g(\vec{\alpha}\times\vec{F}_{AA'BB'})
\end{equation} \\
which, as can be easily verified, is indeed the case for $\vec F_{AA'BB'}$ given by (\ref{covariant4}).
The above development lead us to the following Lagrangian with interaction term \\
\begin{equation} \label{covariant6}
L=\frac{1}{m}\dot{\eta}^{A}_{a}\pi_{aA}
-\frac{g}{m^{2}}\left[(\vec{\tau}^{*}\cdot\vec{\chi}^{AB})\right]_{ab}
\eta_{bB}\pi_{aA}
\end{equation} \\
The Hamiltonian is then given by \\
\begin{equation} \label{covariant7}
H=\frac{\partial L}{\partial \dot{\eta}^{A}_{a}}\dot{\eta}^{A}_{a}- L=
\frac{g}{m^{2}}\left[(\vec{\tau}^{*}\cdot\vec{\chi}_{AB})\right]_{ab}
\eta^{B}_{b}\pi^{A}_{a}
\end{equation} \\
and equals the interaction term. Once expanded, the term $(\vec{\tau}^{*}\cdot\vec{\chi}_{AB})$ is \\
\begin{displaymath}
\left(\begin{array}{cc}
\chi^{3}_{AB} & \chi^{1}_{AB}+i\chi^{2}_{AB} \\
\chi^{1}_{AB}-i\chi^{2}_{AB} & -\chi^{3}_{AB} \\
\end{array}\right)\equiv 
\left(\begin{array}{cc}
\chi^{3}_{AB} & \sqrt{2}\chi^{-}_{AB} \\
\sqrt{2}\chi^{+}_{AB} & -\chi^{3}_{AB} \\
\end{array}\right)
\end{displaymath} \\
with \\
\begin{displaymath} 
\chi^{\bold{\pm}}_{AB}=\frac{1}{\sqrt{2}}(\chi^{1}_{AB}\mp i\chi^{2}_{AB})
\end{displaymath} \\
so, for a typical interaction term, we have \\
\begin{displaymath}
(\vec{\tau}^{*}\cdot\vec{\chi}_{AB})_{ab}\pi^{A}_{a}\eta^{B}_{b}=\chi^{3}_{AB}\pi^{A}_{1}\eta^{B}_{1}+
\sqrt{2}\chi^{-}_{AB}\pi^{A}_{2}\eta^{B}_{1}+\chi^{+}_{AB}\pi^{A}_{1}\eta^{B}_{2}-
\chi^{3}_{AB}\pi^{A}_{2}\eta^{B}_{2}.
\end{displaymath} \\
From (\ref{covariant6}), the associated spinor equations of motion for $\eta_{aA}$ and $\pi_{aA}$ respectively are found to be \\
\begin{displaymath}
-\frac{1}{m}\dot{\pi}_{aA}=\frac{\partial H}{\partial \eta_{a}^{A}}=\frac{g}{m^{2}}(\vec{\tau}^{*}\cdot\vec{\chi}_{BA})_{ba}\pi_{b}^{B}=
\frac{g}{m^{2}}(\vec{\tau}\cdot\vec{\chi}_{BA})_{ab}\pi_{b}^{B}
\end{displaymath} 
\begin{displaymath}
\dot{\eta}_{aA}=-m\frac{\partial H}{\partial\pi_{a}^{A}}=
-\frac{g}{m}(\vec{\tau}^{*}\cdot\vec{\chi}_{AB})_{ab}\eta_{b}^{B}
\end{displaymath} \\
equivalent to \\
\begin{equation} \label{covariant8}
\dot{\eta}_{aA}=-\frac{g}{m}(\vec{\tau}^{*}\cdot\vec{\chi}_{AB})_{ab}\eta^{B}_{b}, 
\end{equation} 
\begin{equation} \label{covariant9}
\dot{\pi}_{aA}=-\frac{g}{m}(\vec{\tau}\cdot\vec{\chi}_{AB})_{ab}\pi^{B}_{b}. 
\end{equation} \\
Once expanded, they become \\
\begin{displaymath}
\begin{array}{c}
\dot{\eta}^{1}_{A}=-\frac{g}{m}\left[\chi^{3}_{AB}\eta^{1B}+\sqrt{2}\chi^{-}_{AB}\eta^{2B}\right] \\
\dot{\eta}^{2}_{A}=-\frac{g}{m}\left[\sqrt{2}\chi^{+}_{AB}\eta^{2B}-\chi^{3}_{AB}\eta^{1B}\right] \\
\dot{\pi}^{1}_{A}=-\frac{g}{m}\left[\chi^{3}_{AB}\pi^{1B}+\sqrt{2}\chi^{+}_{AB}\pi^{2B}\right] \\
\dot{\pi}^{2}_{A}=-\frac{g}{m}\left[\sqrt{2}\chi^{-}_{AB}\pi^{2B}-\chi^{3}_{AB}\pi^{1B}\right]. \\
\end{array}
\end{displaymath} \\
Note that, if instead of $SU(2)$ we consider the symmetry group $U(1)$ (one single infinitesimal 
generator equal to unity and, therefore, one single associated spinorial field $\phi^{AB}$) then, the 
equations of motion of section $\bold{II}$ for the electromagnetic field are recovered.   
Finally, it is convenient to point out that equations (\ref{covariant8}),(\ref{covariant9}) give account
of a coupling 
between the orbital $\{\pi^{A},\eta^{A}\}$ degrees of freedom and
the internal $\{\alpha,\beta\}$ ones.
\section{COLOR QUARK DYNAMICS}
Going further, we would like to extend the above development to $SU(3)$ local phase transformations. If we assume,
for simplicity, a single flavour of quark, the analysis of the preceeding section can be easily extended to any other
symmetry group and, in particular, to describe color quark dynamics in classical Weyl-spinor language. 
This is done just by substituying the $SU(2)$-infinitesimal generators by those of $SU(3)$ in the 
expression for the lagrangian (\ref{covariant5}), so we here shall quote the principal results only.
For a single quark of mass $m$ existing in three color charges, the Lagrangian is ($a=1,2,3$) \\
\begin{equation} \label{covariant10}
L= \frac{1}{m}\dot{\eta}^{A}_{a}\pi_{aA}-
\frac{g_{s}}{m^{2}}[\lambda^{*}_{\bold{q}} W^{\bold{q}}_{AB}]_{ab}\eta^{B}_{b}\pi^{A}_{a},
\end{equation} \\
where $\lambda^{*}_{\bold{q}}$ ($\bold{q}=1,2,...,8$) are the transposed Gell-Mann $SU(3)$ matrices and 
${W}^{\bold{q}}_{AB}$ the $8$ boson spinor fields. The Hamiltonian in this case is given by \\
\begin{equation} \label{covariant11}
H=\frac{g_{s}}{m^{2}}[\lambda^{*}_{\bold{q}} W^{\bold{q}}_{AB}]_{ab}\eta^{B}_{b}\pi^{A}_{a},
\end{equation} \\
and the gauge fields transform as ($\bold{i},\bold{j},\bold{k}=1,2,...,8$) \\
\begin{displaymath}
W'^{\bold{i}}_{AB}=W^{\bold{i}}_{AB}-im\dot{\alpha}^{\bold{i}}\epsilon_{AB}-f^{\bold{ijk}}\alpha^{\bold{j}}(\tau)W^{\bold{k}}_{AB}
\end{displaymath} 
\begin{displaymath}
\bar{W}'^{\bold{i}}_{A'B'}=\bar{W}^{\bold{i}}_{A'B'}+im\dot{\alpha}^{\bold{i}}\epsilon_{A'B'}-f^{\bold{ijk}}\alpha^{\bold{j}}
(\tau)\bar{W}^{\bold{k}}_{A'B'}
\end{displaymath} \\
where $f^{\bold{ijk}}$ are the structure constants of $SU(3)$. For the fourth-rank spinor field one obtains \\
\begin{equation} \label{covariant12}
F^{\bold{i}}_{AA'BB'}=W^{\bold{i}}_{AB}\epsilon_{A'B'}+\bar{W}^{\bold{i}}_{A'B'}\epsilon_{AB}
\end{equation} \\
and $F_{AA'BB'}$ transforms as \\
\begin{displaymath}
F'^{\bold{i}}_{AA'BB'}=F^{\bold{i}}_{AA'BB'}-g_{S}\left[f^{\bold{ijk}}\alpha^{\bold{j}}(\tau)F^{\bold{k}}_{AA'BB'}\right].
\end{displaymath} \\
The equations of motion are now given by \\
\begin{equation} \label{covariant13}
\dot{\eta}_{aA}=-\frac{g_{S}}{m}(\lambda^{*}_{\bold{q}} W^{\bold{q}}_{AB})_{ab}\eta^{B}_{b},
\end{equation} 
\begin{equation} \label{covariant14}
\dot{\pi}_{aA}=-\frac{g_{S}}{m}(\lambda_{\bold{q}}W^{\bold{q}}_{AB})_{ab}\pi^{B}_{b}.
\end{equation} \\
In a more explicit form, these equations are \\
\begin{equation} \label{covariant15}
\left(\begin{array}{c}
\dot{\eta}^{1}_{A} \\
\dot{\eta}^{2}_{A} \\
\dot{\eta}^{3}_{A} \\
\end{array}\right)=-\frac{g_{S}}{m}\left(\begin{array}{ccc}
W^{3}_{AB}+\frac{W^{8}_{AB}}{\sqrt{3}} & W^{1}_{AB}+iW^{2}_{AB} & W^{4}_{AB}+iW^{5}_{AB} \\
W^{1}_{AB}-iW^{2}_{AB} & -W^{3}_{AB}+\frac{W^{8}_{AB}}{\sqrt{3}} & W^{6}_{AB}+iW^{7}_{AB} \\
W^{4}_{AB}-iW^{5}_{AB} & W^{6}_{AB}-iW^{7}_{AB} & -\frac{2W^{8}_{AB}}{\sqrt{3}} \\
\end{array}\right)\left(\begin{array}{c}
\eta^{1B} \\
\eta^{2B} \\
\eta^{3B} \\
\end{array}\right).
\end{equation} \\
\begin{equation} \label{covariant16}
\left(\begin{array}{c}
\dot{\pi}^{1}_{A} \\
\dot{\pi}^{2}_{A} \\
\dot{\pi}^{3}_{A} \\
\end{array}\right)=-\frac{g_{S}}{m}\left(\begin{array}{ccc}
W^{3}_{AB}+\frac{W^{8}_{AB}}{\sqrt{3}} & W^{1}_{AB}-iW^{2}_{AB} & W^{4}_{AB}-iW^{5}_{AB} \\
W^{1}_{AB}+iW^{2}_{AB} & -W^{3}_{AB}+\frac{W^{8}_{AB}}{\sqrt{3}} & W^{6}_{AB}-iW^{7}_{AB} \\
W^{4}_{AB}+iW^{5}_{AB} & W^{6}_{AB}+iW^{7}_{AB} & -\frac{2W^{8}_{AB}}{\sqrt{3}} \\
\end{array}\right)\left(\begin{array}{c}
\pi^{1B} \\
\pi^{2B} \\
\pi^{3B} \\
\end{array}\right).
\end{equation} \\
\section*{Final Remarks}
As already emphasized the Weyl-spinor approach permits an extension of purely
classical physics far beyond electrodynamics. On the other hand, it seems now
that the spinor $master$ $equations$ obtained in paper I can be applied to a 
wide variety of situations and, perhaps (of special interest), in the cosmological
scenario of the primordial quark-gluon plasma state ending in quark confinement.
It is curious to realize that equations (\ref{covariant15}), (\ref{covariant16}), which describe 
the dynamical behaviour of quarks in the presence of gluon fields, mimic the
Lorentz-force of electrodynamics (although with a much greater level of complexity).
This has in fact some experimental support, since it has been already pointed out
the similarity between the energy levels of charmonium ($c\bar{c}$) (due to 
strong forces) and positronium ($e^{+}e^{-}$) (due to electromagnetic forces)
\cite{aitchison}. The assertion made in paper I about the fully generality
of the $master$ $equations$ is then supported by these results.


\begin{thebibliography}{99}

\bibitem{buitrago} J. Buitrago and S. Hajjawi \textit{Spinor extended Lorentz-force-like equation as 
a Consequence of a Spinorial Structure of Space-Time}, J.Math.Phys., $\bold{48}$ $022902$ ($2007$).
%
\bibitem{penrose} R. Penrose and W. Rindler \textit{Spinors and Space-Time}, 
Cambridge Monographs in Mathematical Physics, Vol. $1$, Cambridge Universtiy Press, Cambridge, 
England ($1984$/$1986$).
%
\bibitem{aitchison} I. Aitchison and A. Hey \textit{Gauge Theories in Particle Physics },
Institute of Physics Publishing, Vol. $1$, $150$ South Independence Mall West, Philadelphia, PA $19106$, USA ($2003$).
%
\end{thebibliography}
\end{document}